\documentclass[english,aps,10pt,english,aps,twocolumn,showpacs,preprintnumbers,amsmath,amssymb,superscriptaddress,groupedaddress]{revtex4}
\pdfoutput=1
\usepackage[latin9]{inputenc}
\usepackage{graphicx}  
\usepackage{float}
\usepackage{amsmath}
\usepackage{amssymb}
\usepackage{graphicx}
\usepackage{esint}
\usepackage{babel}
\usepackage{hyperref}
\usepackage{placeins}
\usepackage{color}
\usepackage[dvipsnames]{xcolor}

\begin{document}

\title{Optomechanical Detection of Light with Orbital Angular Momentum}

\author{Hamidreza Kaviani}
\affiliation{National Institute for Nanotechnology, Edmonton, AB, T6G 2M9, Canada}
\affiliation{Institute for Quantum Science and Technology, University of Calgary, Calgary, AB, T2N 1N4, Canada}
\author{Roohollah Ghobadi}
\affiliation{Institute for Quantum Science and Technology, University of Calgary, Calgary, AB, T2N 1N4, Canada}
\author{Bishnupada Behera}
\affiliation{National Institute for Nanotechnology, Edmonton, AB, T6G 2M9, Canada}
\affiliation{Institute for Quantum Science and Technology, University of Calgary, Calgary, AB, T2N 1N4, Canada}
\author{Marcelo Wu}
\affiliation{Physical Measurement Laboratory, National Institute of Standards and Technology, Gaithersburg, MD 20899, USA}
\author{Aaron Hryciw}
\affiliation{National Institute for Nanotechnology, Edmonton, AB, T6G 2M9, Canada}
\author{Sonny Vo}
\author{David Fattal}
\affiliation{Leia Inc., 2440 Sand Hill Rd, STE 303 Menlo Park, CA 94025, USA}
\author{Paul Barclay}
\email{pbarclay@ucalgary.ca}
\affiliation{National Institute for Nanotechnology, Edmonton, AB, T6G 2M9, Canada}
\affiliation{Institute for Quantum Science and Technology, University of Calgary, Calgary, AB, T2N 1N4, Canada}

\date{Compiled \today}

\begin{abstract}
We present an optomechanical device designed to allow optical transduction of orbital angular momentum of light.  An optically induced twist imparted on the device by light is detected using an integrated cavity optomechanical system based on a nanobeam slot-mode photonic crystal cavity. This device could allow measurement of the orbital angular momentum of light when photons are absorbed by the mechanical element, or detection of the presence of photons when they are scattered into new orbital angular momentum states by a sub-wavelength grating patterned on the device. Such a system allows detection of a $l = 1$ orbital angular momentum field with an average power of $3.9\times10^3$ photons modulated at the mechanical resonance frequency of the device and can be extended to higher order orbital angular momentum states. 
\end{abstract}

\maketitle

\section{Introduction}

It is well known that photons have linear momentum \cite{ref:lebedew1901eel} and spin angular momentum in the form of circular polarization \cite{ref:beth1936mdm}. However, only recently has it been appreciated that photons also possess orbital angular momentum (OAM), following Allen et al.'s discovery that light with a helical wavefront has non--zero OAM \cite{ref:allen1992oam}. Since then, a vast range of applications for OAM of light have been proposed. These range from high bandwidth data transfer \cite{ref:wang2012tfd, bozinovic2013toa} and quantum cryptography \cite{ref:vallone2014fqk, ref:mirhosseini2015hqc}, to optical tweezers for biological applications \cite{ref:oneil2002ien}. Light with OAM has been generated using pitchfork holograms \cite{ref:bazhenov1990lbw, ref:heckenberg1992gop}, spiral phase plates \cite{ref:sueda2004lgb,ref:leach2004ovs}, Dove prisms \cite{ref:leach2002moa}, cylindrical lens mode converters \cite{ref:allen1992oam,ref:beijersbergen1993alm}, liquid crystals $q$-plates \cite{ref:karimi2009egs}, metasurfaces such as sub-wavelength gratings \cite{ref:biener2002fhb,ref:vo2014swg}, plasmonic nano-antennas \cite{ref:karimi2014goo, ref:yu2011lpp}, optical phased arrays \cite{ref:sun2014gip} and microrings \cite{ref:cai2012icp}. Similarly, methods for measuring the OAM of the light have seen rapid development, including techniques based on forked diffraction gratings \cite{ref:mair2001eoa}, interferometry \cite{ref:leach2004imm}, apertures \cite{ref:sztul2006dil, ref:Zhou2014dms} and image reformatting \cite{ref:berkhout2010eso}. OAM of light can also be measured using the torque that it exerts on incident objects. Mechanical detection of spin angular momentum was first demonstrated by Beth \cite{ref:beth1936mdm} and the direct mechanical measurement of the OAM of light has been achieved using torsional pendulums \cite{beijersbergen2005moa} and microscopic particles \cite{he1995dot}. However, mechanical detection of the OAM of light using on-chip devices has not been achieved to date. Such devices would enable OAM to be used as a fingerprint in quantum non-demolition measurement of photons \cite{ref:braginsky1980qnm}, as well as provide a platform for interfacing free-space OAM carrying optical fields to nanophotonic components.

The orbital angular momentum $\hbar l$ of a photon is determined by the OAM quantum number $l$ describing its helical wavefront and is in principle unbounded \cite{ref:allen1992oam}. As a result, OAM can be much larger than spin angular momentum which is limited to $\pm\hbar$. Although from a practical point of view efficient generation and measurement of large OAM is challenging, light beams with up to $l=300$ have been demonstrated \cite{ref:fickler2012qeh}. The torque per photon due to a change in OAM of light upon its interaction with an optical element is given by $\tau = \hbar \Delta l /\Delta t$, where $\Delta l$ is the change of the OAM of light and $\Delta t$ is the time duration of the photon pulse. For continuous optical excitation, we can express the torque in terms of the power of the incident light: 
\begin{equation}
\tau (t)=\frac{\Delta l P(t)}{\omega}.
\label{eq:torque_p}
\end{equation}  
where $\omega$ is the frequency of the incident light. In order to measure this change in OAM via $\tau$, it is necessary to create a system that modifies $l$, and whose mechanical motion can be both efficiently actuated by $\tau$ and sensitively monitored.

Here we propose and analyze an optomechanical photonic crystal cavity to detect the torque exerted by light on a photonic nanostructure, allowing measurement of the OAM of light, as well as non-absorbing optical field detection via light's OAM degree of freedom. Optomechanical photonic crystal cavities localize light to sub-wavelength volumes where it interacts strongly with nanomechanical resonances of the device, resulting in coupling between nanomechanical motion and the optical cavity resonance frequency and linewidth. Large optomechanical coupling and high-quality factor ($Q_o$) optical resonances possible in these devices provide sensitive transduction of mechanical motion via a change in the cavity's optical response. This enables ultrasensitive measurement of sources of force \cite{ref:sun2012fdc,ref:gavartin2012aho}. For example, optomechanical cavities have been used to realize accelerometers \cite{ref:krause2012ahm} and integrated atomic force detection systems \cite{ref:liu2012wcs}. Although  optomechanical nanoscale torque sensing \cite{ref:wu2014ddo,ref:kim2013nto} has recently been studied within the context of torque magnetometry \cite{ref:wu2017not,ref:kim2017maf} and spin detection of photons \cite{ref:he2016omp}, OAM detection has not previously been explored using a nanophotonic cavity optomechanics platform. 

\begin{figure*}
\centering
\includegraphics[width=2 \columnwidth]{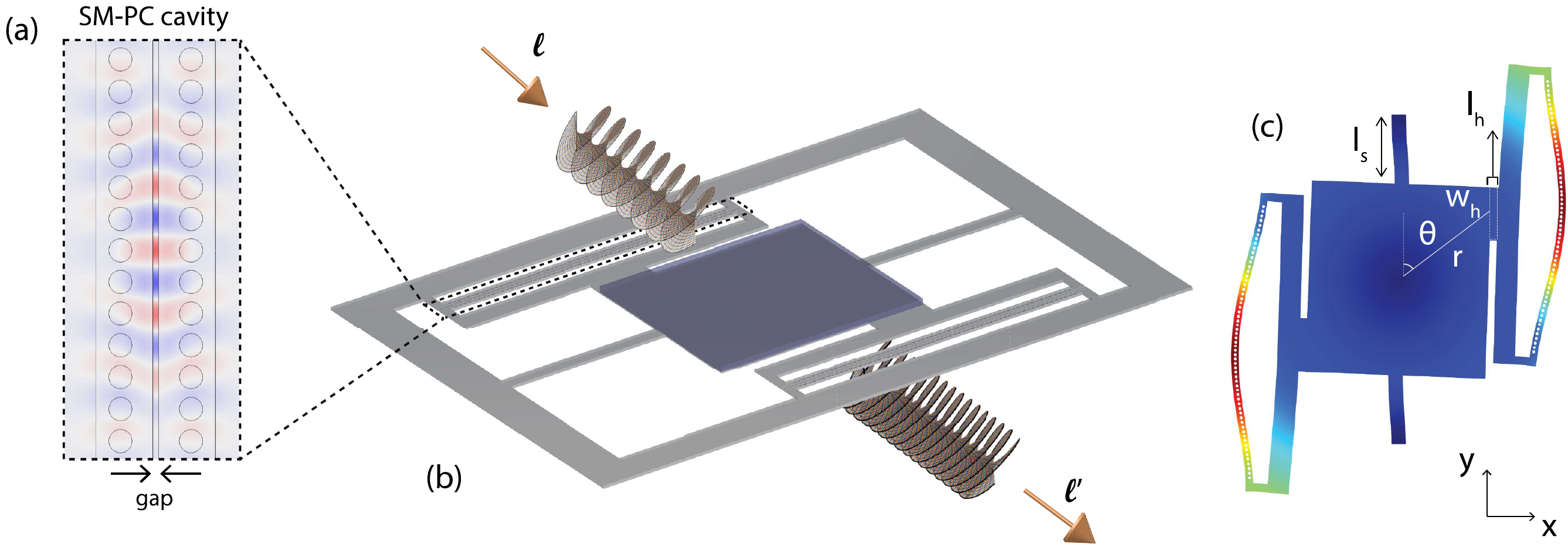}
\caption{Overview of the device geometry and elements. (a) Slot-mode photonic crystal cavity fundamental mode electric field ($x$ component) distribution. (b) Isometric view of the OAM detector. A helical light beam illuminates the square pad which is suspended by thin supports. Depending on the structure of additional material attached to the central square pad (blue), light exerts a torque on the pad due to its change in OAM during reflection, refraction, absorption or transmission. The cavity from (a) is attached to the square pad by a hanger whose dimensions $w_h$ and $l_h$ are indicated in (c). The pad motion is coupled to the nanobeam as shown by the typical displacement profile shown in (c). For a given frequency of actuation, this motion is due to both motions of the pad and excitation of the nanobeam mechanical modes. Optical readout of the cavity via using a fiber taper waveguide allows sensitive measurement of this motion.}
\label{fig:schematics}
\end{figure*}

Figure \ref{fig:schematics} shows a schematic of the OAM detection device proposed and studied here. It consists of a central suspended pad connected to a nanobeam slot-mode photonic crystal optomechanical cavity. This cavity shares some characteristics with photonic crystal zipper cavities \cite{ref:eichenfield2009oc}. However, unlike the zipper cavity, the slot-mode photonic crystal cavity supports optical modes concentrated in the gap between two nanobeams, as shown by the electric field profile for the cavity's fundamental mode in Fig.\ \ref{fig:schematics}(a). This ``air--band'' cavity mode, which is analogous to slot-modes of two-dimensional photonic crystal from in Ref.\ \cite{ref:sun2012sos}, has high sensitivity to motion of the nanobeam that changes the slot gap width, and as a result, changes the optical response of the cavity through a dispersive optomechanical coupling. As shown in Fig.\ \ref{fig:schematics}(b), one of the cavity's nanobeams is attached to the central pad, while the other is fixed to the surrounding chip. Optical actuation of the central pad by OAM can shift the nanobeam's centre of mass position through ``twisting'' of the central pad, and through excitation of the mechanical ``bouncing'' mode of the nanobeam. Both of these types of displacement can be seen in the simulated displacement profile of the device shown in Fig.\ \ref{fig:schematics}(c), and are discussed in more detail below.  The cavity optical mode properties can be monitored using a fiber taper optical waveguide evanescently coupled to the cavity.

The device presented here is designed from a 370 nm thick silicon nitride (SiN) layer. SiN is chosen because of its low mechanical \cite{ref:verbridge2008mnr} and optical \cite{ref:barclay2006ifc} loss, and its high tensile stress, which makes it suitable for large scale suspended devices with mechanical quality factors up to $Q_m=10^8$ \cite{ref:krause2012ahm, ref:reinhardt2016ust, ref:norte2016mrq}. 
The slot width separating the two nanobeams is 100 nm, and finite difference time domain simulations were used to calculate the field profile of the cavity's fundamental slot-mode shown in Fig.\ \ref{fig:schematics}(a).  This mode is predicted to have a wavelength $\lambda = 1428$ nm, optical quality factor $Q_o >10^6$, and mode volume $V \sim 0.19 \ (\lambda/n_\text{SiN})^3$ (defined by the peak field strength). Additional details of this device design are provided in \cite{kaviani2019ods}. Although in the following analysis of this system we will calculate the optomechanical coupling to the hybridized mechanical modes of the full device, the cavity's baseline optomechanical performance is quantified by its optomechanical coupling coefficient of $g_{OM}/2\pi>64$ GHz/nm to the fundamental mechanical bouncing mode of the nanobeams. The effective mass of this mode is $m_{eff}=27$ pg, and is small compared to the slot-mode cavity in Ref.\ \cite{ref:sun2012sos} due to its one-dimensional nature. For calculation of $V$, $g_{OM}$, and $m_{eff}$, we have used the definitions in Ref.\ \cite{ref:eichenfield2009mdc} and COMSOL finite element simulations.

The OAM of light incident onto the central pad of the device can be converted to torque in several ways. If the torsional pad is coated with an absorptive layer and illuminated by helical light, the OAM of light changes from $l$ to $l^{\prime}=0$. On the other hand, if the torsional pad is patterned with a suitably engineered metasurface, a helical beam with OAM number $l$ can be changed during transmission or reflection to, in principle, arbitrary OAM number $l^{\prime}$ defined by the metasurface geometry. Following this interaction, as shown in Fig.\ \ref{fig:schematics}(c), the motion of the torsional pad is mechanically coupled to the motion of the slot-mode cavity nanobeam via a rectangular hanger with length and width of $l_h$ and $w_h$, respectively.

\section{Optomechanical device properties}

\begin{figure}
\centering
\includegraphics[width=\columnwidth]{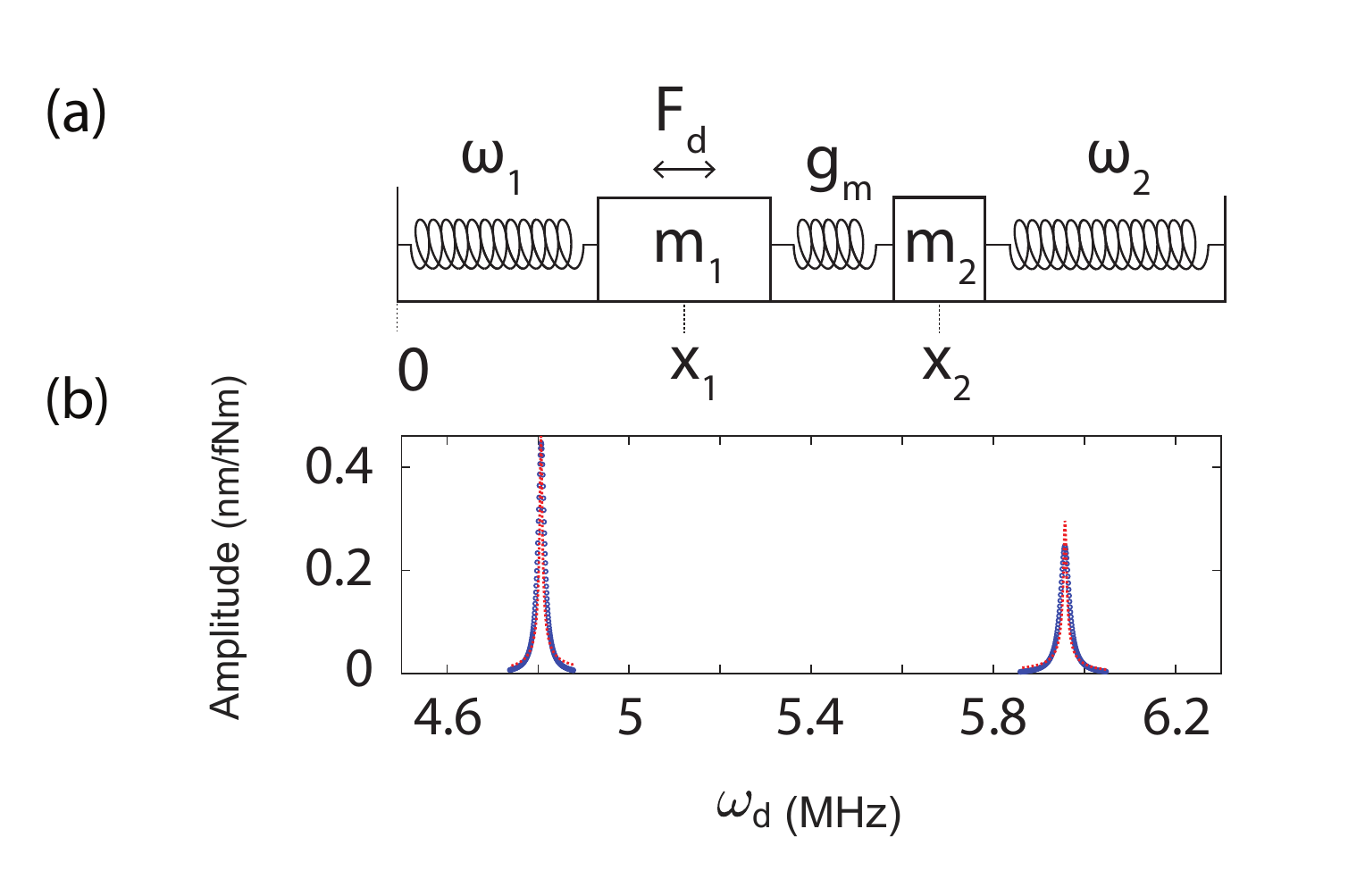}
\caption{(a) Schematic of the coupled oscillator model describing the interaction of the nanobeam bouncing mode and the central pad twisting mode. (b) Simulated displacement $x_2$ of the nanobeam as a function of the frequency of a torsional drive applied to the central pad edges, for device parameters $l_s=12\,\mu$m , $w_h=7\,\mu$m, $l_h=1\,\mu$m. Blue dots are simulated points, and the red dashed lines are fits from the coupled oscillator model.}\label{fig:two_oscillators}
\end{figure}

To study the interaction between torsional excitation of the central pad and both the nanobeam's center of mass position and its vibrational motion, we performed numerical simulations and developed a semi-analytic coupled harmonic oscillator model. Figure \ref{fig:two_oscillators}(a) shows a cartoon representation of this model, where the central  pad and the nanobeam are represented as oscillators with masses and natural frequencies of $m_1$, $\omega_1$ and $m_2$, $\omega_2$, respectively. The coupling between the torsional pad and the nanobeam is represented by a spring with natural frequency of $g_m$. Variables $x_1$ and $x_2$ are the maximum displacement from equilibrium of the torsional pad and nanobeam in the $\hat{x}$ direction, respectively. Note that $x_1$ is related to radius $r$ and angle of rotation $\theta$ defined in Fig.\ \ref{fig:schematics}(c) by $x_1=r\cos\theta\Delta\theta$. 
The equations of motion for this system are:
\begin{subequations}
\label{eq:equations_of_motion}
\begin{align}
 \ddot{x_1} &= -\omega_1^2 x_1 - \gamma_1 \dot{x_1}+ \sqrt{\frac{m_2}{m_1}}g_m^2 x_2 + \frac{F_d}{m_1} e^{- i \omega_d t},\\
 \ddot{x_2} &= -\omega_2^2 x_2 - \gamma_2 \dot{x_2}+ \sqrt{\frac{m_1}{m_2}}g_m^2 x_1, \
\end{align}
\end{subequations}
where $\gamma_1$, $\gamma_2$ are mechanical damping rates and $F_d$ and $\omega_d$ are the amplitude and frequency of an external drive force applied to the central pad. Fourier transforming Eqs.\ \ref{eq:equations_of_motion}, we can solve for $x_2$, 
\begin{equation}
x_2(\omega) = \frac{g_m^2}{\sqrt{m_1 m_2} ((\chi_1(\omega) \chi_2(\omega))^{-1}- g_m^4)} F_d(\omega-\omega_d),
\label{eq:x2}
\end{equation}
where $\chi_{{1,2}}(\omega)$ are mechanical susceptibilities $\chi_{1,2}(\omega)=(\omega_{1,2}^2 - \omega^2 - i \gamma_{1,2} \omega)^{-1}$.

\begin{figure}
\centering
\includegraphics[width=\columnwidth]{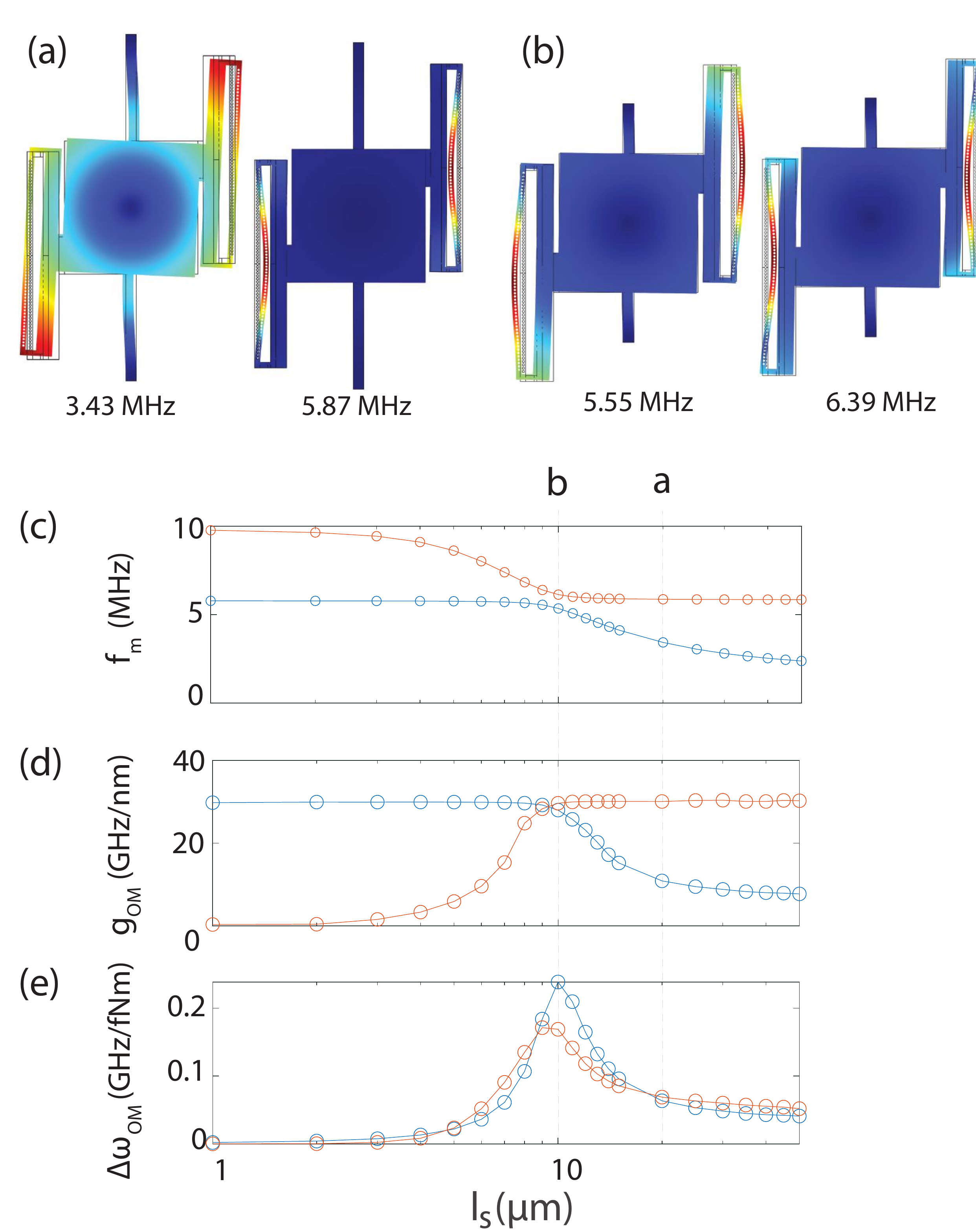}
\caption{ Displacement profiles of the hybridized central pad twisting mode and nanobeam bouncing modes for (a) $l_s = 20\,\mu\text{m}$ and (b) $l_s = 10\,\mu\text{m}$ as labeled in (c). Dependence of hybridized mode (c) frequency, (d) optomechanical coupling coefficient, and (e) torsional optomechanical frequency shift. }
\label{fig:two_modes}
\end{figure}

The validity of this model can be evaluated using finite element method software (COMSOL) to calculate the mechanical and optomechanical properties of our OAM detection device. Figure \ref{fig:two_oscillators}(b) shows the result of a simulation of the maximum displacement of the nanobeam when we drive the central pad by applying a torque at varying frequency $\omega_d$. This torque has been implemented in the simulation by applying tangential forces to the sides of the torsional pad. An initial stress of $S_0 = 1~\text{GPa}$ in the SiN layer was included in the simulations. Predictions from the coupled oscillator model described by Eq.\ \ref{eq:x2} are shown by a red dotted line and fit the finite element simulation results well with $g_m$ as a fitting parameter. Note that for these simulations, $\gamma_{1,2}$ corresponding to mechanical quality factor $Q_m = 500$ are fixed in COMSOL.  

The peaks in Fig.\ \ref{fig:two_oscillators}(b) correspond to resonant excitation of two mechanical modes of the device: the twisting mode of the central pad, and the bouncing mode of the nanobeam, which for the central pad support length chosen for this simulation have widely separated frequencies of $\omega_1/2\pi = 4.81$ MHz and $\omega_2/2\pi = 5.96$ MHz, respectively. The twisting mode peak is stronger as it is driven directly, while the bouncing mode is driven indirectly through the nanobeam's mechanical coupling to the central pad. However, Eq.\ \ref{eq:x2} predicts that $x_2$ can be enhanced if the oscillators are tuned near resonance ($\omega_1=\omega_2$). This can be achieved in our design by tuning the central pad twisting mode frequency via changing the support length $l_s$ labeled in Fig.\ \ref{fig:schematics}(c).  

Displacement profiles of the mechanical modes obtained from COMSOL simulations are shown in Figs.\ \ref{fig:two_modes}(a) and (b)  for different values of $l_s$. Figure \ \ref{fig:two_modes}(c) shows the simulated $l_s$ dependence of their resonance frequencies. The nanobeam bouncing mode frequency does not depend on $l_s$, while the frequency of the central pad twisting mode decreases with increasing $l_s$. For values of $l_s$ where the modes are not on resonance, their mode profiles are dominantly twisting-- or bouncing--like, as shown in Fig.\ \ref{fig:two_modes}(a). Near $l_s = 10~\mu\text{m}$ they are on-resonance, and an avoided crossing is observed due to modal coupling. This coupling is evident in the modes' mechanical displacement profiles when $l_s$ is tuned to the center of the anti-crossing: as shown in Fig.\ \ref{fig:two_modes}(b) the twisting and bouncing modes are hybridized into even and odd combinations. The degree of splitting between their frequencies is related to the mode coupling $g_m$. 

The efficacy with which these modes can be used to detect OAM strongly depends on their optomechanical coupling coefficient $g_{OM}$. We calculate $g_{OM}$ using the perturbation theory discussed in Ref.\ \cite{ref:johnson2002ptm, ref:eichenfield2009oc} input with the simulated displacement profiles of the mechanical modes for the full device and the field profile of the optical cavity's fundamental slot-mode. The resulting $g_{OM}$ of each mechanical mode for varying $l_s$ is shown in Figure \ref{fig:two_modes}(d). We see that when the modes are not resonant, the nanobeam bouncing mode has nearly constant $g_{OM} \sim 32\,\text{GHz/nm}$, while the central pad's twisting mode's $g_{OM}$ is much lower. When the modes are tuned on-resonance, they have equal $g_{OM}$, as expected given their hybridized nature.

\begin{figure}
\centering
\includegraphics[width=\columnwidth]{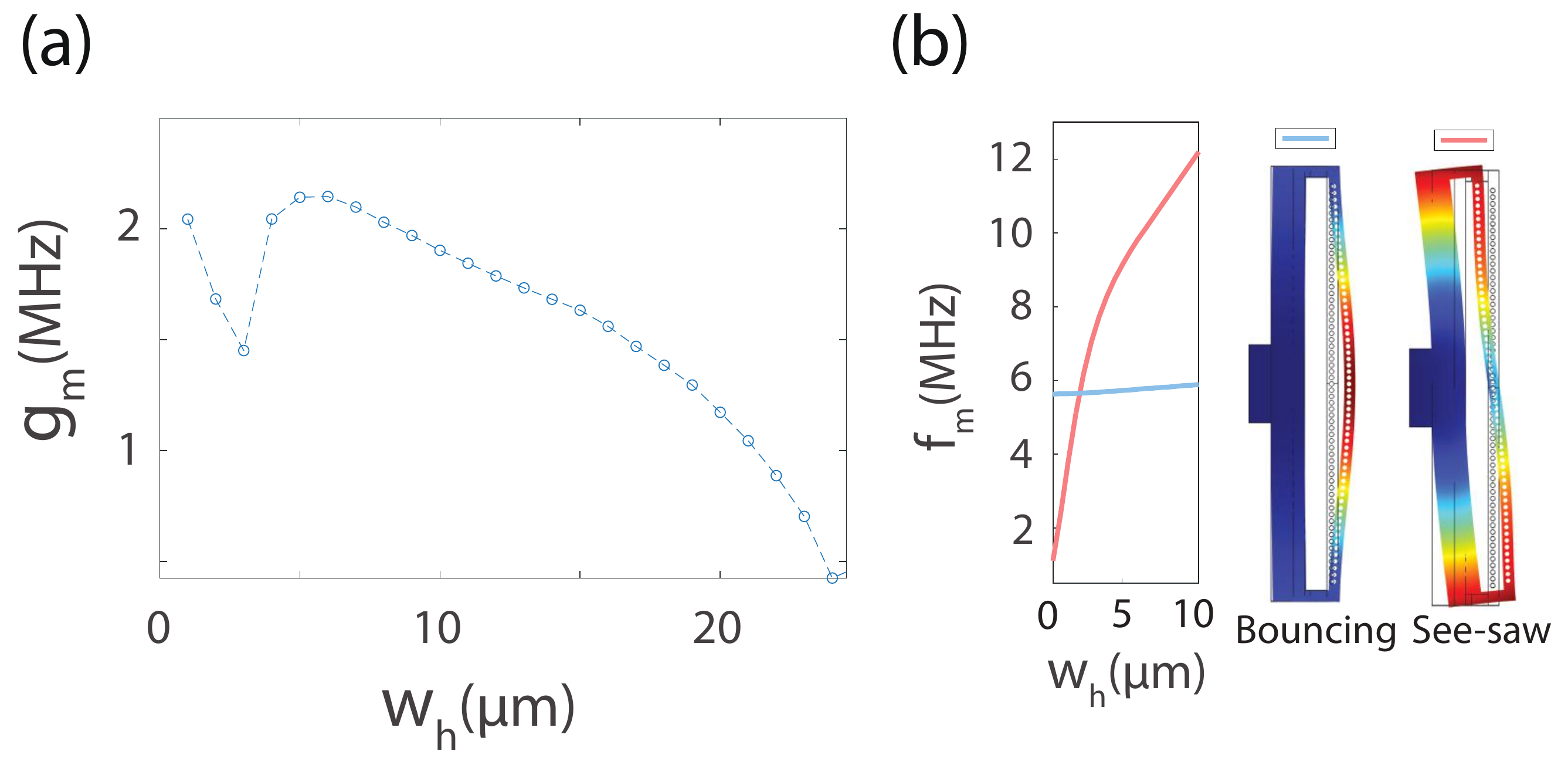}
\caption{(a) Coupling $g_{m}$ between the central pad twisting mode and the nanobeam beam bouncing mode for varying hanger width, extracted from COMSOL simulations of the hybridized mechanical mode frequencies. (b) Dependence of the mechanical frequencies of the nanobeam bouncing (blue) and see-saw (red) resonances on $w_h$.}
\label{fig:gm}
\end{figure}

We further characterize the optomechanical properties of the device by calculating the optomechanical shift of the cavity mode frequency, $\Delta\omega_{OM}$, per $\text{fN}\cdot\text{m}$ of torque applied to the central pad at the calculated mechanical mode frequencies shown in Fig.\ \ref{fig:two_modes}(c). As shown in Fig.\ \ref{fig:two_modes}(e), when $l_s$ is set to tune the mechanical modes are on--resonance, $\Delta\omega_{OM}$ peaks due to excitation of the nanobeam bouncing mode by the twisting motion.  The enhancement to $\Delta\omega_{OM}$ when the central pad twisting and the nanobeam bouncing modes are resonant is related to the resonator coupling $g_m$, as shown in Eq.\ \ref{eq:x2}. We study the dependence of this coupling on the geometry of the hanger connecting the resonators in Fig.\ \ref{fig:gm}(a), which shows $g_m$ for varying  $w_h$, as calculated by fitting the anti-crossing of $\omega_m(l_s)$ for each $w_h$  to the coupled mode model  with $g_m$ as a fitting parameter. We find that the mechanical coupling decreases with increasing $w_h$. Intuitively, this is due to the fact that as the width of hanger increases, the hanger becomes more centered on the pad ($\theta=90^{\circ}$) and the average horizontal center of mass motion of the nanobeam becomes negligible. For small values of hanger width, the see-saw mode of the nanobeam, whose frequency and mode profile are shown in Fig.\ \ref{fig:gm}(b), becomes close to resonance with the twisting and bouncing mode frequencies. Its coupling rate to the central pad's twisting mode is much higher than that of nanobeam bouncing mode, reducing the amount of energy coupled to the nanobeam bouncing mode. Therefore, we chose $w_h=7\,\mu$m for the operation of our OAM sensor in the remainder of the analysis.    

 
 \section{Torque sensitivity}

To predict the sensitivity of this device to OAM generated torque, we need to consider sources of noise, including thermal Brownian motion ($\tau_{th}$), photon shot noise ($\tau_{SN}$), detector noise ($\tau_{DN}$) and back-action noise ($\tau_{BA}$). These noise-equivalent torques combine to determine the minimum detectable torque of the device: $\tau_{min}=\sqrt{(\tau_{th})^{2}+(\tau_{SN})^{2}+(\tau_{DN})^{2}+(\tau_{BA})^{2}}$. Applied torque in our system is related to $x_{max}$ by
\begin{equation}
x_{max}(\omega)=[ m_{eff}r_{eff} (\omega_m^2-\omega^2+\frac{i\omega\omega_m}{Q_m})]^{-1}\tau(\omega),
\label{eq:xtau}
\end{equation}
where $r_{eff}$ is the effective lever arm which depends on the position of the force and maximum displacement with respect to the axis of rotation of the system. The lower $r_{eff}$, the lower the moment of inertia is, which leads to a higher sensitivity to external torques. Due to the relatively complex geometry of our device we extract $r_{eff}$ by fitting Eq.\ \ref{eq:xtau} to the finite element simulated $x_{max}$ when an external torque is applied to the central pad.

Using Eq.\ \ref{eq:xtau} and the fluctuation dissipation theorem, the thermal noise equivalent torque is:
\begin{equation}
\tau_{th}(\omega)=\sqrt{\frac{4k_BT\omega_mm_{eff}r_{eff}^2 }{Q_m}}.
\label{eq:tau_th}
\end{equation}
Equation \ref{eq:tau_th} suggests that to achieve higher sensitivities, a low effective mass, a small effective lever arm and small mechanical frequency are desirable, as well as low environment temperature, $T$, and high mechanical quality-factor. 
\begin{figure}
\centering
\includegraphics[width=\columnwidth]{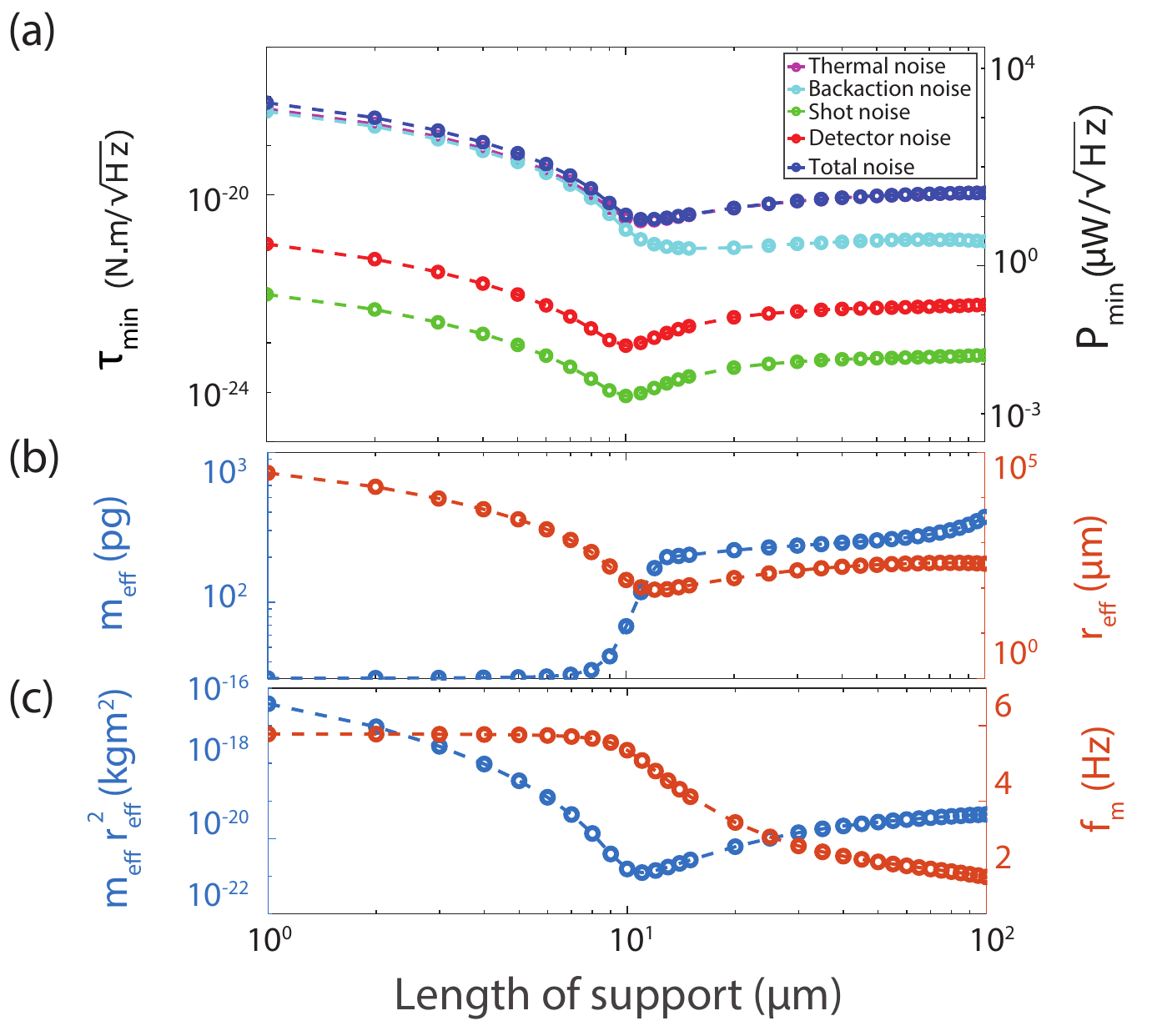}
\caption{(a) The minimum detectable torque (left axis) and minimum detectable optical incident power (right axis) as a function of support length, with contribution from different noise sources shown. (b) Effective mass (left) and effective lever arm (right) as a function of support length. (c) Contributing factors to the thermal noise as a function of length of supports.}
\label{fig:noises}
\end{figure} 

The importance of optomechanical device parameters $Q_o$ and $g_{OM}$ for reaching a regime of OAM detection limited by thermal noise is revealed by expressions for the torque equivalent photon shot noise and detector noise. The shot noise optical power spectral density is given by $S_{P}^{SN}=2 \hbar \omega_{0} P_{det}/\eta_{qe}$ \cite{ref:fox2006qoa}, and the corresponding torque equivalent noise is 
\begin{equation}
\tau_{SN}(\omega_m)=\frac{m_{eff} \omega_m^2 r_{eff}\sqrt{S_{PP}^{SN}}}{\mid \frac{dT}{d\Delta} \mid Q_m P_{det} g_{OM}}, 
\label{eq:tau_SN}
\end{equation}
where $\frac{dT}{d\Delta}$ is the optical wavelength dependent slope of the transmission profile of the fiber coupled optical cavity mode, and $P_{det}$ is the measured power at the detector. The electronic noise of a typical photoreceiver (Newport 1811) used for directly monitoring intensity fluctuations of the fiber taper output is $P_{DN}=2.5$ pW/$\sqrt{\text{Hz}}$, with corresponding torque equivalent noise given by
\begin{equation}
\tau_{DN}(\omega_m)=\frac{m_{eff} \omega_m^2 r_{eff}P_{DN}}{\mid \frac{dT}{d\Delta} \mid Q_m P_{det} g_{OM}}.  
\label{eq:tau_DN}
\end{equation}
Lastly, we analyze optomechanical backaction noise resulting from radiation pressure fluctuations of the power coupled from the readout laser into the photonic crystal cavity. The force per photon in this cavity is given by $\hbar g_{OM}$, and the torque equivalent noise associated with photon number uncertainty in the cavity is given by
\begin{equation}
\tau_{BA}(\omega_m)=2\hbar g_{OM} r_{eff} \sqrt{\frac{n_{\text{cav}}}{\kappa}},
\label{eq:tau_BA}
\end{equation} 
where $n_{cav}$ is intracavity photon number and $\kappa$ is the cavity decay rate. 


Combining these noise sources, we can predict the minimum detectable torque of the device. This is shown in Fig.\ \ref{fig:noises}(a), where we have assumed that the measurement and source of torque is at frequency $\omega_m$ resonant with the lower frequency branch of the hybridized device modes from Fig.\ \ref{fig:two_modes}(c). From this, we see that $\tau_{min}$ minimizes near $l_s \sim 10\,\mu\text{m}$ where the central pad twisting and nanobeam bouncing modes are on resonance with another. This behavior is a result of the different $l_s$ dependence of $r_{eff}$ for nanobeam bouncing versus central pad twisting modes, as shown in Fig.\ \ref{fig:noises}(b) together with the behavior of $m_{eff}$ in these two regimes. The corresponding effective moment of inertia is shown in Fig.\ \ref{fig:noises}(c). These calculation were performed assuming $Q_{o}=10^{6}$, $Q_{m}=10^{6}$, cryogenic temperature (4 K) operation, $P_{det}=0.1 \,\mu\text{W}$, $\Delta l=1$ and hanger geometry $w_{h}=7\,\mu$m and $l_{h}=1\,\mu$m. From this we see that $\tau_{min}\approx 3.22\times 10^{-21}\,$N$\cdot$m/$\sqrt{\text{Hz}}$ is expected to be achievable.

\section{Transmissive OAM detection}

\begin{figure}
\centering
\includegraphics[width=\columnwidth]{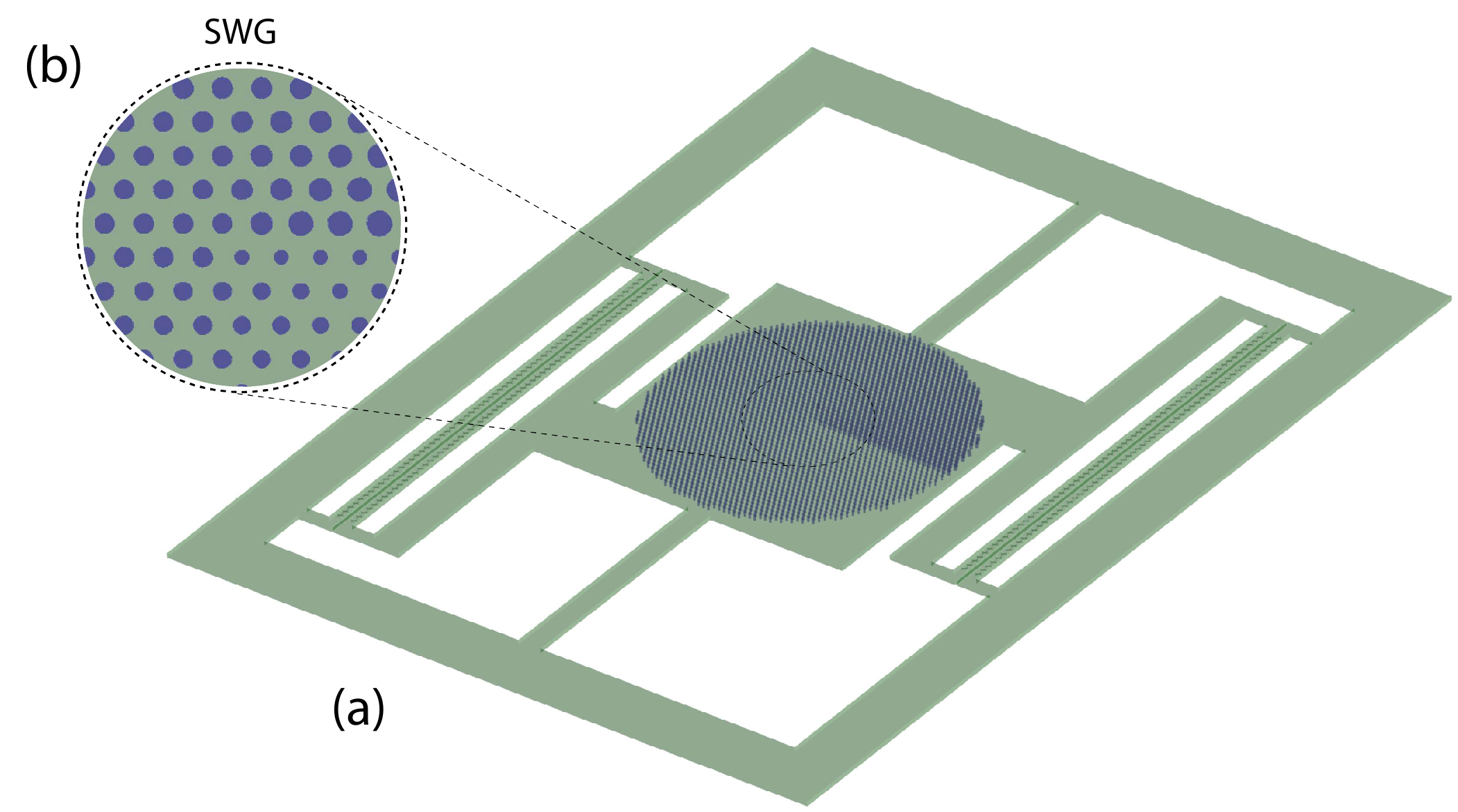}
\caption{(a) Transmissive OAM detection system. a-Si (blue) pillars have been patterned on top of the SiN (green) optomechanical device in order to form an SWG. (b) Top view picture of the SWG designed for $\Delta l=1$, showing the azimuthally varying pillar diameters.}
\label{fig:SWG_Schematics}
\end{figure}

We next focus on the possibility of detecting OAM of an optical field by converting it to torque via its transmission through the central pad. This would allow detection of light without destroying its linear momentum and intensity, which could be used for a non-demolition photon detection scheme. To achieve this, we designed a sub-wavelength grating (SWG) shown in Fig.\ \ref{fig:SWG_Schematics}, whose on-chip chiral pattern converts OAM to mechanical torque. This SWG is a high refractive index contrast pattern \cite{ref:mateus2004umu} composed of amorphous silicon (a-Si) pillar structures with refractive index n$=3.62$ \cite{ref:vo2014swg} patterned on the SiN central pad. High refractive index contrast gratings can operate off-resonance (reflective) \cite{ref:lu2010phn,ref:fattal2010fdg} and on-resonance (transmissive) \cite{ref:vo2014swg}, and in both cases have demonstrated ultra-broadband operation and high capability of modulating wavefront phase. In our design, the transmissive SWG has a $20\,\mu$m diameter and consists of $450\,\text{nm}$ thick a-Si pillars. The pillars are arranged in a hexagonal lattice with lattice constant $\Lambda=360\, \text{nm}$. Varying the duty cycle as a function of angle, by changing pillars' diameter from $110$ nm to $210$ nm in increments of $10$ nm, results in modulation of the local effective refractive index and corresponding spatially varying phase shift of light transmitted through the SWG that depends on the azimuthal angle.  This allows the incident beam OAM state to be changed by $\Delta l$ determined by the SWG design. The physical principle of this SWG is similar to that discussed in \cite{ref:fattal2012lds, ref:vo2014swg}.

\begin{figure}
\centering
\includegraphics[width=\columnwidth]{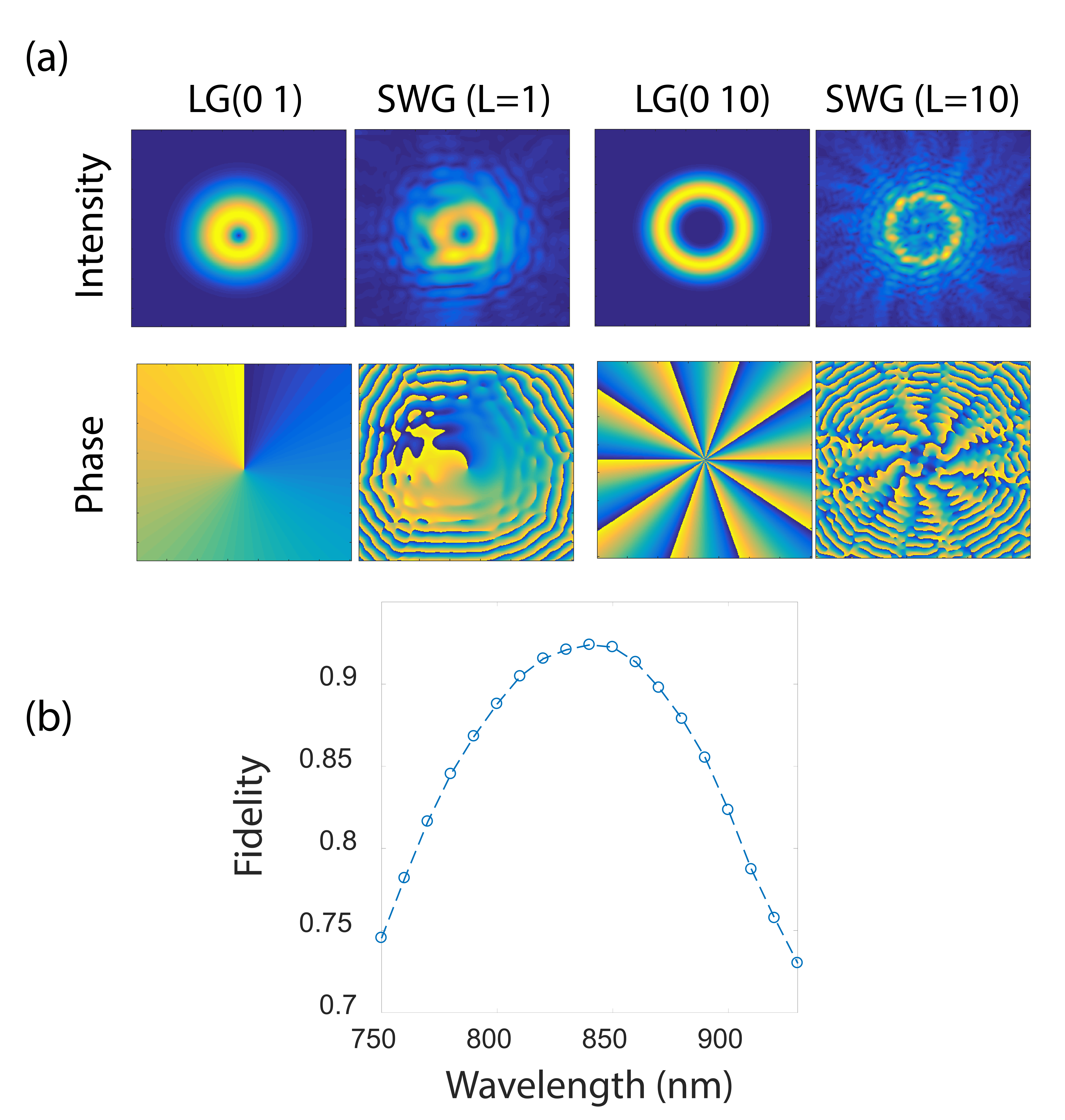}
\caption{(a) FDTD simulated transmitted optical field intensity and phase profiles of SWGs of type $\Delta l=1$ and $\Delta l=10$ for a Gaussian incident field. Also shown are the ideal optical field intensity and phase profiles of LG modes with $l = 1$ and $l = 10$. (b) Fidelity of conversion of a Gaussian beam transmitted through the $\Delta l=1$ SWG to the $l = 1$ LG mode as a function of the incident light wavelength.}
\label{fig:OAM}
\end{figure}

Figure \ref{fig:OAM}(a) shows the simulated optical phase and intensity profiles of a Gaussian beam transmitted through this SWG when it is designed for two cases: $\Delta l=1$ and $\Delta l=10$. These results were obtained using Lumerical finite-difference time-domain (FDTD) software. Comparing these results to the analytically calculated phase and intensity profiles of  Laguerre-Gaussian (LG) beams with $l=1$ and $l=10$, also shown in Fig.\ \ref{fig:OAM}(a), we see that the SWGs convert the input Gaussian beam ($l=0$) to a LG beam with $l=1$ or $l=10$, depending on the grating design.  Deviations between the ideal profiles and the simulated transmitted profiles are also visible in Fig.\ \ref{fig:OAM}(a), and indicate that the SWG conversion efficiency is not ideal. This efficiency can be benchmarked by considering the fidelity ($F$) of output compared to a perfect LG beam, and SWG transmission ($T_{SWG}$) \cite{ref:liu2016ocg}. This yields an OAM conversion efficiency given by $F \times T_{SWG}=0.90\times0.92=0.83$ for the $\Delta l=1$ SWG operating at a wavelength of 840 nm. Using this conversion efficiency, together with Eq.\ \eqref{eq:torque_p} and assuming an input beam modulated at $\omega_m$, we predict the minimum detectable optical power scattered from $l =0 $ to $l = 1$ to be $8.7\,\mu\text{W}/\sqrt{\text{Hz}}$, as shown as a function of $l_s$ in Fig.\ \ref{fig:noises}(a). The sensitivity of the device linearly scales with the change in OAM and can reach less than a $1\,\mu\text{W}/\sqrt{\text{Hz}}$ using the SWG designed for $\Delta l=10$. This assumes the same operating conditions and device parameters as the torque sensitivity analysis above, and that the optical field is modulated with unity contrast at the frequency of the lower branch of the device's hybridized mechanical resonances.

\section{Pulsed operation}

\begin{figure}
\centering
\includegraphics[width=\columnwidth]{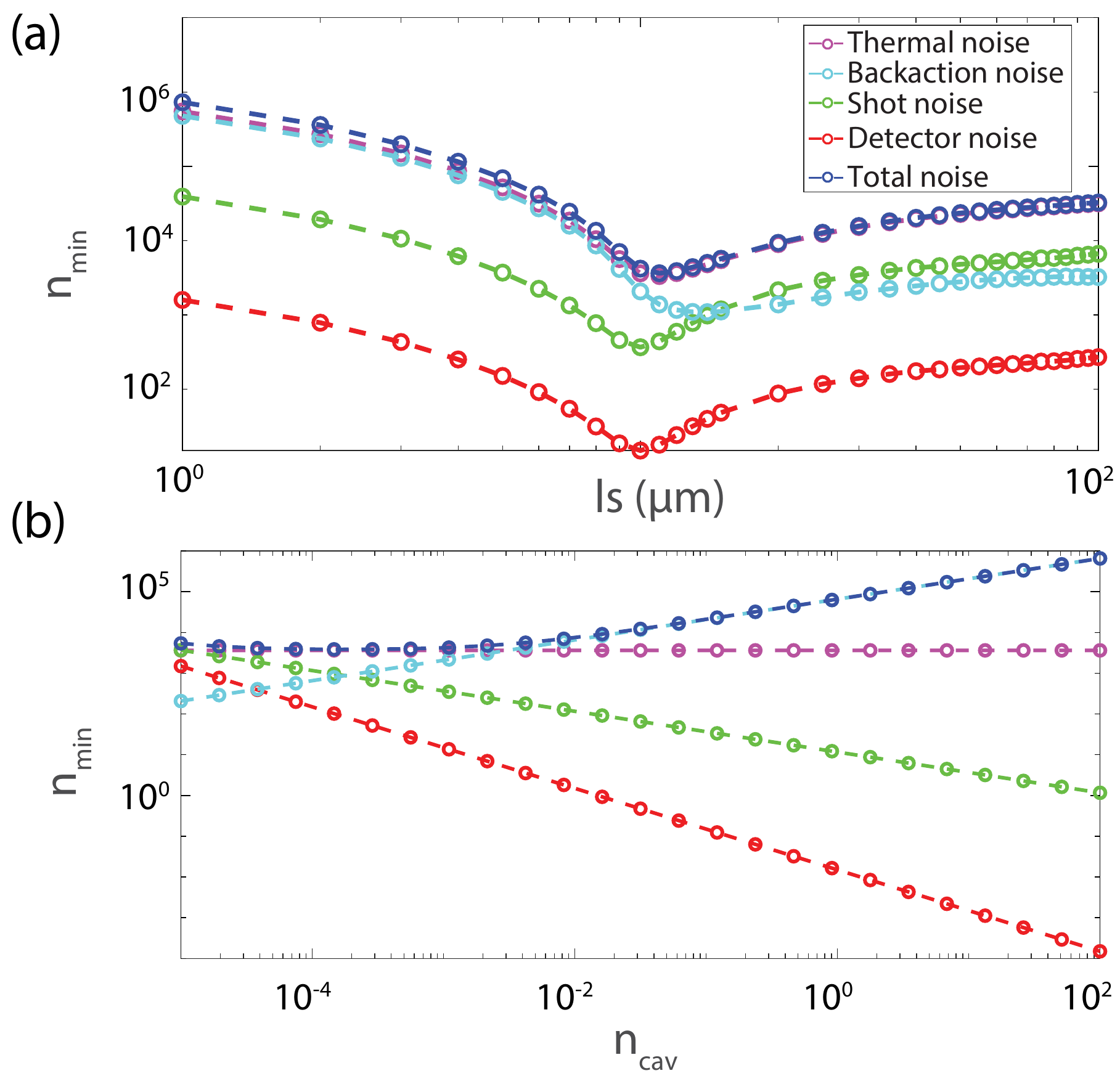}
\caption{(a) Minimum detectable photon number per pulse with repetition rate on mechanical resonance, as a function of support length. Device parameters and operating conditions: $Q_o=10^6$, $Q_m=10^8$, $\Delta l$=10 and $T=10\,$mK. (b) Minimum detectable photon number per pulse as a function of readout interactivity photon number for support length $l_s=10\,\mu$m.}
\label{fig:noise_photon_number}
\end{figure}

Next, we study the sensitivity of our device when we drive it with optical pulses instead of a harmonically modulated continuous wave. The SWG used above has a wide optical bandwidth of operation, as shown by the wavelength dependence of its fidelity shown in Fig.\ \ref{fig:OAM}(b), allowing short optical pulses to drive the device. The optical power of a pulse train whose pulse width $\Delta t$ is much smaller than the repetition time $1/f_r$ can be expressed as $P(t)=\sum_{m=-\infty}^{+\infty}n\hbar\omega_{c}f_{r} e^{i2 \pi m f_{r} t}$ where $n$ and $\omega_{c}$  are photon number per pulse and the carrier frequency of the incident light, respectively. This shows that in order to resonantly drive the mechanical device, the pulse repetition rate should be set to mechanical frequency ($2\pi f_r=\omega_m$). In this case, the optical drive power of the pulse train is $P(\omega_m)=n\hbar\omega_{c}f_r$. Using $P(\omega_m)$ and Eq.\ \ref{eq:torque_p}, the exerted torque by the pulse train can be calculated. Figure \ref{fig:noise_photon_number}(a) shows the predicted minimum detectable number of photons $n_{min}$ per pulse as a function of $l_s$, indicating that pulses with as few as $3.9\times10^3$ photons can be detected. In plotting Fig.\ \ref{fig:noise_photon_number}(a) we assume the ideal but achievable conditions of $Q_o=10^6$, $Q_m=10^8$ \cite{ref:norte2016mrq,ref:reinhardt2016ust,ref:tsaturyan2017unr,ref:ghadimi2018ese}, $\Delta l=10$, $n_{\text{cav}}=10^{-3}$ and $T=10\,$mk. We also assume detector noise $P_{DN}=3.8\times 10^{-17}$W$/\sqrt{\text{Hz}}$, which is reachable by using a superconducting nanowire single-photon detector, for example Single Quantum Eos detector. \cite{ref:singlequantum}

As can be seen in Fig.\ \ref{fig:noise_photon_number}(a), the measurement sensitivity is maximized when $l_s$ is chosen to tune the central pad twisting mode and the nanobeam bouncing mode onto resonance. To study the impact of backaction from the readout laser, which is significant for the idealized device parameters assumed here, in Fig.\ \ref{fig:noise_photon_number}(b) we show $n_\text{min}$ as a function of intracavity photon numbers $n_{\text{cav}}$ for $l_s=10\,\mu$m (when torsional pad and nanobeam bouncing modes are in resonance). This shows that $n_{\text{min}}=3.9\times 10^3$ is the optimal operating condition for these device parameters.     

\section{Refractive and Absorptive detection}
The device studied here also allows OAM detection in refractive or absorptive schemes without the need of patterning an SWG on the central pad. The OAM change, and resulting torque, when a photon is absorbed by the central pad is given by the total OAM of the input field, and is independent of the central pad geometry. However, this scheme is destructive as the photons are destroyed upon detection. Implementation of absorptive detection with the device studied here would require operation at shorter wavelengths below the transparency window of SiN, utilization of multiphoton or impurity related absorption processes, or modification of the central pad material to enhance its optical absorption.

In the case of a refractive detection scheme, which can be realized with a planar surface, the change in OAM is given by $\Delta l = 0.5(\cos\theta_i/\cos\theta_r+\cos\theta_r/\cos\theta_i)l$ where $\theta_i$ and $\theta_r$ are incident and refracted angles \cite{ref:fedoseyev2008toa}. In this scheme, the magnitude of an incident field's OAM and linear momentum are conserved while their directions are changed.      

\section{Conclusion} 

In conclusion, we have designed an optomechanical system that enables non-destructive measurement of light via the torque induced by its OAM on the device. The device has a torque sensitivity of $\tau_{\text{min}}=3.22\times 10^{-21}\,$N$\cdot$m/$\sqrt{\text{Hz}}$, allowing OAM detection of optical fields with a sensitivity of $P_{\text{min}}=8.7\,\mu\text{W}/\sqrt{\text{Hz}}$, assuming $Q_{o}=10^{6}$, $Q_{m}=10^{6}$ and cryogenic temperature $T=4\,$K. Considering the existing state-of-the-art performance of similar SiN devices, detection of $3.9\times 10^3$ photons in a single pulse is achievable. This number could be further reduced by using SWGs with higher order OAM conversion ($\Delta l >10$) and designing a photonic crystal nanobeam cavity with a lower mechanical frequency.

\section{Acknowledgments}

The authors would like to thank Ebrahim Karimi and Boris Braverman for fruitful discussions. This work was supported by the National Research Council, Canada (NRC), Alberta Innovates, the National Sciences and Engineering Research Council of Canada (NSERC), and the Canada Foundation for Innovation (CFI).

\bibliography{nano_bib}

\end{document}